\documentclass[twocolumn,aps,floatfix]{revtex4}
\usepackage{graphicx}
\usepackage{dcolumn}
\usepackage{bm}
\usepackage{mhchem}

\begin{document}

\title{High $p_{t}$ squeezed-out n/p ratio as a probe of $K_{\rm{sym}}$ of the symmetry energy}

\author{Ya-Fei Guo}
\author{Gao-Chan Yong}\email{yonggaochan@impcas.ac.cn}

\affiliation{Institute of Modern Physics, Chinese Academy of Sciences, Lanzhou 730000, China\\
School of Nuclear Science and Technology, University of Chinese Academy of Sciences, Beijing 100049, China}

\begin{abstract}

By involving the constraints of the slope of nuclear symmetry energy $L$ into the question of  determination of the high-density symmetry energy, one needs to probe the curvature of nuclear symmetry energy $K_{\rm{sym}}$. Based on the Isospin-dependent Boltzmann-Uehling-Uhlenbeck (IBUU) transport model, effects of the curvature of nuclear symmetry energy on the squeezed-out nucleons are demonstrated in the semi-central Au+Au reaction at 400 and 600 MeV/nucleon. It is shown that the squeezed-out isospin-dependent nucleon emissions at high transverse momenta are sensitive to the curvature of nuclear symmetry energy. The curvature of nuclear symmetry energy at saturation density thus can be determined by the high momentum squeezed-out isospin-dependent nucleon emissions experiments from the semi-central Au+Au reaction at 400 or 600 MeV/nucleon.

\end{abstract}

\maketitle

\section{Motivation}

Nowadays the study of the equation of state (EoS) of
nuclear matter is one of the most hot topic in nuclear community \cite{li08,bar05,topic,GW171,GW181,GW182}.
The EoS at density $\rho$ and isospin asymmetry
$\delta$ ($\delta=(\rho_n-\rho_p)/(\rho_n+\rho_p)$) is usually expressed as
\begin{equation}
E(\rho ,\delta )=E(\rho ,0)+E_{\text{sym}}(\rho )\delta ^{2}+\mathcal{O}%
(\delta ^{4}),
\end{equation}%
where $E_{\text{sym}}(\rho)$ is the density-dependent nuclear symmetry energy.
Currently the EoS of isospin symmetric nuclear matter $E(\rho, 0)$ is roughly
determined \cite{pawl2002} but the EoS of neutron-rich matter, especially the high-density symmetry energy, is still very controversial \cite{Guo14}.
The high-density symmetry energy is closely related to the physics of neutron stars \cite{Lat01}, such as their stellar radii and cooling rates \cite{Lat04,Vil04,Ste05}, the gravitational-wave frequency \cite{gwf,gwf2}, the gamma-ray bursts \cite{grb}, the r-process nucleosynthesis \cite{rpn1,rpn2,rpn3} in neutron star mergers \cite{GWth,GW170817}.
Experimentally, probing the high-density symmetry energy by the measurements of pion and nucleon, triton and $^{3}$He yields ratio in isotope Sn+Sn ractions at about 300 MeV/nucleon are being carried out at RIBF/RIKEN in Japan \cite{sep,shan15}. Such projects are also being carried out/planned at FOPI/GSI and CSR/Lanzhou \cite{fopi16,csr}, the Facility for Rare Isotope Beams (FRIB) in the Untied States \cite{frib} and the Rare Isotope Science Project (RISP) in Korea \cite{korea}. And some progress has been made by measuring nucleon and light charged cluster flows at FOPI/GSI \cite{fopi16}.

The density-dependent symmetry energy around saturation can be Taylor expanded in terms of a few bulk parameters \cite{issac2009},
\begin{equation}\label{esym2}
E_{sym}(\rho)=E_{sym}(\rho_0)+L\left(\frac{\rho-\rho_0}{3\rho_0}\right)
+\frac{K_{sym}}{2}\left(\frac{\rho-\rho_0}{3\rho_0}\right)^2,
\end{equation}
where $E_{sym}(\rho_0)$ is the value of the symmetry energy at saturation point and the quantities $L$, $K_{sym}$ are related to its slope and curvature, respectively, at the same density point,
\begin{equation}
\begin{array}{c}
\displaystyle{L=3\rho_0\frac{\partial E_{sym}(\rho)}{\partial \rho} \Big |_{\rho=\rho_0}} \ , \,\,\,\,
\displaystyle{K_{sym}=9\rho_0^2\frac{\partial^2E_{sym}(\rho)}{\partial \rho^2} \Big |_{\rho=\rho_0}}. \end{array}
\end{equation}
Over the last two decades the most probable magnitude $E_{\rm sym}(\rho_0)= 31.7\pm 3.2$ MeV and slope $L= 58.7\pm 28.1 $ MeV of the nuclear symmetry energy at saturation density have been found through surveys of 53 analyses \cite{Oertel17,Li13}. The curvature $K_{\rm{sym}}$ is known probably to be in the range of $-400 \leq K_{\rm{sym}} \leq 100$ MeV \cite{Tews17,Zhang17}.
In fact, there is an interplay between $L$ and $K_{\rm{sym}}$ terms in Eq.~(\ref{esym2}). In case the slope $L$ is fixed by some suitable constraints, then
the curvature $K_{\rm{sym}}$ (which is more related to the high-density symmetry energy) can be nicely constrained. Since nowadays the values $E_{\rm sym}(\rho_0)$ and the slope $L$ are constrained in narrow ranges, one must involve these progress in determination of the high-density symmetry energy. That is to say, the question of the determination of the high-density symmetry energy should be timely converted into the constraints of the coefficient $K_{\rm{sym}}$.

Besides the strong correlation between the neutron-star tidal deformability (which related to the gravitational waves in neutron star merger \cite{GW171,GW181,GW182}) and $K_{\rm{sym}}$ \cite{tsangns}, the curvature of nuclear symmetry energy is also closely related to the incompressibility coefficient of nuclear matter, which is difficult to determine by properties of neutron-rich nuclei thus remains an important open problem \cite{issac2009,pie2009}.

To probe the curvature of nuclear symmetry energy at saturation density, based on the updated Isospin-dependent Boltzmann-Uehling-Uhlenbeck (IBUU) transport model, we studied the isospin-dependent nucleon elliptic flow and the squeezed-out free neutron to proton ratio at high transverse momenta in the semi-central Au+Au reaction at 400 and 600 MeV/nucleon and find they are both sensitive to the curvature of nuclear symmetry energy at saturation density.
At 600 MeV incident beam energy, the squeezed-out free neutron to proton ratio at high transverse momenta is more sensitive to the curvature of nuclear symmetry energy than at 400 MeV beam energy while the isospin-dependent nucleon elliptic flow is less sensitive to the curvature of nuclear symmetry energy when changing incident beam energy from 400 to 600 MeV/nucleon.
The curvature of nuclear symmetry energy at saturation density thus can be directly probed by the high $p_{t}$ squeezed-out n/p in heavy-ion collisions.

\section{model and methods}

The updated Isospin-dependent
Boltzmann-Uehling-Uhlenbeck (IBUU) transport model originates from the IBUU04 model \cite{lyz05}.  In this model, effects of the short-range-correlations are appropriately taken into account \cite{yong20171,yong20172}.
The neutron and proton density distributions in nucleus are given by the Skyrme-Hartree-Fock with Skyrme M$^{\ast}$ force parameters \cite{skyrme86}. In initialization of colliding nuclei, the proton and neutron momentum distributions with high-momentum tails are employed \cite{yong20171,yong20172,sci08,sci14,yongcut2017}.
The isospin- and momentum-dependent single nucleon mean-field
potential reads
\begin{eqnarray}
U(\rho,\delta,\vec{p},\tau)&=&A_u(x)\frac{\rho_{\tau'}}{\rho_0}+A_l(x)\frac{\rho_{\tau}}{\rho_0}\nonumber\\
& &+B\Big(\frac{\rho}{\rho_0}\Big)^{\sigma}(1-x\delta^2)-8x\tau\frac{B}{\sigma+1}\frac{\rho^{\sigma-1}}{\rho_0^\sigma}\delta\rho_{\tau'}\nonumber\\
& &+\frac{2C_{\tau,\tau}}{\rho_0}\int
d^3\,p'\frac{f_\tau(\vec{r},\vec{p^{'}})}{1+(\vec{p}-\vec{p^{'}})^2/\Lambda^2}\nonumber\\
& &+\frac{2C_{\tau,\tau'}}{\rho_0}\int
d^3\,p'\frac{f_{\tau'}(\vec{r},\vec{p^{'}})}{1+(\vec{p}-\vec{p^{'}})^2/\Lambda^2},
\label{buupotential}
\end{eqnarray}
where $\rho_0$ is the saturation density, $\tau, \tau'$=1/2(-1/2) is for neutron (proton).
$\delta=(\rho_n-\rho_p)/(\rho_n+\rho_p)$ is the isospin asymmetry,
and $\rho_n$, $\rho_p$ denote neutron and proton densities,
respectively. The parameter values $A_u(x)$ = 33.037 - 125.34$x$
MeV, $A_l(x)$ = -166.963 + 125.34$x$ MeV, B = 141.96 MeV,
$C_{\tau,\tau}$ = 18.177 MeV, $C_{\tau,\tau'}$ = -178.365 MeV, $\sigma =
1.265$, and $\Lambda = 630.24$ MeV/c. Note here that in the single particle potential, the parameter $x$ can change with density (while the empirical values of nuclear matter at normal density are kept) to mimic different density-dependent symmetry energy.
And in the model the isospin-dependent baryon-baryon scattering cross section in medium is reduced compared with their free-space value. More details on the present used model can be found in Ref.~\cite{yong20171}.

\begin{figure}[t]
\centering
\includegraphics[width=0.5\textwidth]{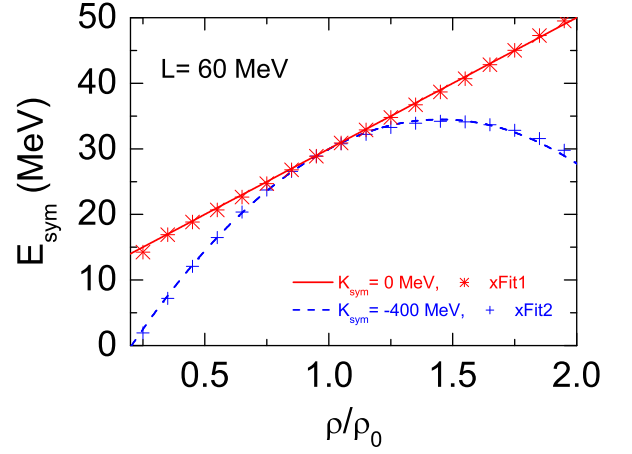}
\caption{(Color online) Density-dependent symmetry energies for curvatures $K_{sym}= 0$ and $K_{sym}= -400$ MeV derived from Eq.~(\ref{buupotential}) (symbols) and Eq.~(\ref{esym2}) (lines) with $E_{\rm sym}(\rho_0)= 30 $ MeV and slope $L= 60$ MeV. } \label{esym}
\end{figure}
To probe the curvature $K_{sym}$ of neutron-rich matter at saturation density and keep the parameters $E_{\rm sym}(\rho_0)$ and slope $L$ fixed, we in Eq.~(\ref{buupotential}) use two density-dependent $x$ parameters to plot two density-dependent symmetry energies by fixing $E_{\rm sym}(\rho_0)= 30$ MeV and $L= 60$ MeV but letting curvatures $K_{sym}= 0$ and $K_{sym}= -400$ MeV, respectively.
For $K_{sym}= 0$ case,
\begin{eqnarray}
x&=& -0.3688\Big(\frac{\rho}{\rho_{0}}\Big)^{5}+ 3.1516\Big(\frac{\rho}{\rho_{0}}\Big)^{4}\nonumber\\
& &- 10.379\Big(\frac{\rho}{\rho_{0}}\Big)^{3}+
16.59\Big(\frac{\rho}{\rho_{0}}\Big)^{2}\nonumber\\
& &- 13.39\Big(\frac{\rho}{\rho_{0}}\Big)+4.9229.
\end{eqnarray}
For $K_{sym}= -400$ case,
\begin{eqnarray}
x&=& 0.2374\Big(\frac{\rho}{\rho_{0}}\Big)^{3}-
1.1231\Big(\frac{\rho}{\rho_{0}}\Big)^{2}\nonumber\\
& &+ 2.0456\Big(\frac{\rho}{\rho_{0}}\Big)-0.539.
\end{eqnarray}
With these settings, employing the density and momentum dependent single particle potential Eq.~(\ref{buupotential}), the density-dependent symmetry energies (shown with symbols) for curvatures $K_{sym}= 0$ and $K_{sym}= -400$ MeV with $E_{\rm sym}(\rho_0)= 30 $ MeV and $L= 60$ MeV are shown in Fig.~\ref{esym}. One can see that, in the density range $0.2-2\rho_{0}$, the symmetry energies derived from Eq.~(\ref{buupotential}) fit the symmetry energies (shown with lines) derived from Eq.~(\ref{esym2}) with $E_{\rm sym}(\rho_0)= 30$ MeV and $L= 60$ MeV  quite well.

\section{Results and Discussions}

\begin{figure}[t]
\centering
\includegraphics[width=0.5\textwidth]{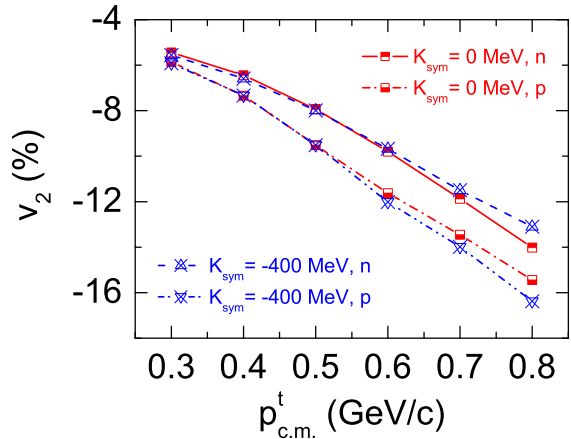}
\caption{(Color online) Neutron and proton elliptic flows $v_{2}$ as a function of transverse momentum in the semi-central Au+Au reaction (b= 7 fm, $|(y/y_{beam})_{c.m.}|\leq$ 0.5) at incident beam energy of 400 MeV/nucleon with $K_{sym}= 0$ and $K_{sym}= -400$ MeV, respectively.} \label{v2}
\end{figure}
Nucleon directed and elliptic flows in
heavy-ion collisions can be derived from the
Fourier expansion of the azimuthal distribution
\cite{Voloshin,dan1998,lip2001,flow,wang14,fan2018}, i.e.,
\begin{equation}
\frac{dN}{d\phi}\propto1+2
\displaystyle{\sum_{i=1}^{n}}v_{n}\cos(n\phi).
\end{equation}
The nucleon elliptic flow $v_{2}$ is obtained from
\begin{equation}
v_{2}=\langle\cos(2\phi)\rangle=\langle\frac{p_{x}^{2}-p_{y}^{2}}{p_{t}^{2}}\rangle
\end{equation}
and can be a potential observable to probe the density-dependent nuclear symmetry energy \cite{lip2001,flow,fan2018}.
Fig.~\ref{v2} shows nucleon elliptic flow as a function of transverse momentum in the semi-central Au+Au reaction at 400 MeV/nucleon with $K_{sym}= 0$ and $K_{sym}= -400$ MeV, respectively. It is first seen that, due to the Coulomb repulsion, the strength of proton elliptic flow is larger than that of neutron. Because the gradient of the high-density symmetry energy with $K_{sym}= 0$ case is larger than that with $K_{sym}= -400$ MeV (as shown Fig.~\ref{esym}) and the symmetry energy is repulsive for neutrons and attractive for protons, the larger effects of the high-density symmetry energy with $K_{sym}= 0$ increase (decrease) the strength of neutron (proton) elliptic flow. It is thus seen that for neutron elliptic flow, the strength is larger for $K_{sym}= 0$ than that for $K_{sym}= -400$ MeV whereas for proton elliptic flow the strength is smaller for $K_{sym}= 0$ that for $K_{sym}= -400$ MeV. The latter is qualitatively consistent with that founding in Ref. \cite{lip2001}.

\begin{figure}[t]
\centering
\includegraphics[width=0.5\textwidth]{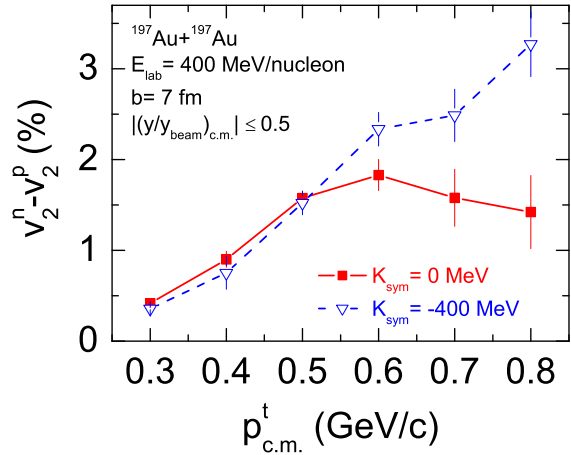}
\caption{(Color online) Difference of neutron and proton elliptic flows $v^{n}_{2}-v^{p}_{2}$ as a function of transverse momentum in the semi-central Au+Au reaction (b= 7 fm, $|(y/y_{beam})_{c.m.}|\leq$ 0.5) at incident beam energy of 400 MeV/nucleon with $K_{sym}= 0$ and $K_{sym}= -400$ MeV, respectively.} \label{v2diff}
\end{figure}
To enlarge the effects of the curvature $K_{sym}$ of the symmetry energy at saturation density, it is good to make a difference of neutron and proton elliptic flows. Fig.~\ref{v2diff} shows the difference of neutron and proton elliptic flows as a function of transverse momentum in the semi-central Au+Au reaction at incident beam energy of 400 MeV/nucleon with $K_{sym}= 0$ and $K_{sym}= -400$ MeV, respectively. It is shown that at lower transverse momenta, the difference of neutron and proton elliptic flows is less sensitive to the curvature $K_{sym}$ of the symmetry energy. At high transverse momenta, however, the effects of the curvature $K_{sym}$ of the symmetry energy on the difference of neutron and proton elliptic flows are quite evident. The difference of neutron and proton elliptic flows with $K_{sym}= -400$ MeV is larger than that with $K_{sym}= 0$. The $v^{n}_{2}-v^{p}_{2}$ at high $p_{t}$ thus can be a potential probe of the curvature $K_{sym}$ of the symmetry energy at saturation density. We also analyzed the $v^{n}_{2}-v^{p}_{2}$ elliptic flow at high $p_{t}$ in the semi-central Au+Au reaction at 600 MeV/nucleon and find that it is not sensitive to the $K_{sym}$, this is consistent with the finding in Ref.~\cite{ditoro10}, which showed that the elliptic flow is not sensitive to the symmetry potential at 600 AMeV if that symmetry potential gives effective masses $m_{n}^{*}>m_{p}^{*}$ \cite{npali05}.

\begin{figure}[t]
\centering
\includegraphics[width=0.5\textwidth]{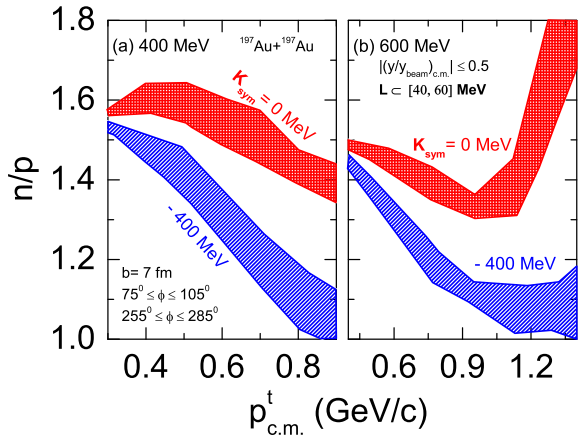}
\caption{(Color online) Effects of the curvature of nuclear symmetry energy at saturation density on the transverse momentum distribution of the
ratio of mid-rapidity ($|(y/y_{beam})_{c.m.}|\leq$ 0.5)) neutrons to protons emitted in the reaction of Au+Au at incident beam energy of 400 (panel (a)) and 600 (panel (b))
MeV/nucleon and impact parameter of b= 7 fm. The azimuthal angle cut is $%
75^{\circ}\leq \phi \leq 105^{\circ}$ and $255^{\circ}\leq \phi
\leq285^{\circ}$, to make sure that the free nucleons are from the
direction perpendicular to the reaction plane. The bandwidths denote uncertainties of the slope $L$ from 60 to 40 MeV \cite{cur2016,l22}.} \label{squeezernp}
\end{figure}
It is known that the squeezed-out nucleons or elliptic flows (emitted in the direction perpendicular to the reaction plane) in semi-central collisions carry more direct information about the high density phase of the reaction \cite{greiner,bert88,cas90,aich91,Reisdorf97,danie02,yong07,ditoro10}. Fig.~\ref{squeezernp} shows the squeezed-out neutron to proton ratio of nucleons emitted in the direction perpendicular to the reaction plane in Au+Au at 400 (left panel) and 600 (right panel) MeV/nucleon. It is clearly seen that the effects of the curvature of the symmetry energy on the squeezed-out neutron to proton ratio n/p are quite evident, especially at high transverse momenta. Since the slope of the high-density symmetry energy with $K_{sym}= 0$ is larger than that with $K_{sym}= -400$ MeV, one sees the values of the squeezed-out n/p with $K_{sym}= 0$ are higher than that with $K_{sym}= -400$ MeV. From both panels, it is seen that, around $p_{t}$ = 0.8 or 1.3 GeV/c, the effects of the curvature of the symmetry energy on the squeezed-out n/p reach about 40-60\%, which are both much larger than the ratio of integrating neutron and proton elliptic flows \cite{cozma}. The squeezed-out n/p thus can be one of the most potential probe of the curvature of the symmetry energy and can be carried out on the rare isotope reactions worldwide.

Since it is not easy to judge which uncertainty combination in the transport model affects the observable much in certain case, here we do not intend to study how the observable is affected by dozens of permutations and combinations of model uncertainties, such as permutations and combinations of several kinds of momentum dependent symmetry potential, several kinds of isospin-dependent in-medium cross section, several kinds of short-range correlation (different fractions of nucleon in the high-momentum tail, different isospin-dependences of neutrons and protons in the high-momentum tail, different forms of momentum distribution, etc.). To reduce the systematic errors and some other theoretical uncertainties to some extend, it is better to experimentally use double ratios in two isotopically different reactions \cite{double1,double2}. In this case the medium-mass isotopic reaction systems, such as $^{132}$Sn+$^{124}$Sn over $^{108}$Sn+$^{112}$Sn seem executable \cite{yong20171}.

Although the effects of the curvature of the symmetry energy on the high $p_{t}$ squeezed-out n/p reach about 50\%, the effects of the slope $L$ on this observable still reach about 7-10\% when changing the slope $L$ from 60 to 40 MeV \cite{cur2016,l22}. At density about $2\rho_{0}$, affections of the slope $L$ on the $K_{sym}$ sensitive observable are inevitable since there is no direct way to differentiate the contributions from $L$ or $K_{sym}$ terms in Eq. (\ref{esym2}). At very high densities, although the $K_{sym}$ term
may contribute more than the $L$ term, it is hard to find a $K_{sym}$ sensitive observable in heavy-ion collisions due to the high temperature of the dense matter formed in energetic heavy-ion collisions.

Since the slope $L$ is more related to the symmetry energy around saturation density than the $K_{sym}$, considering complexity of the transport model, it is better to constrain
the slope $L$ of the symmetry energy at saturation density by using properties of neutron-rich nuclei or matter at ground state \cite{nuc1,nuc2,nuc3,nuc4,nuc5,nuc6,nuc7,nuc8,nuc9,nuc10,nuc11}.
Because $K_{sym}$ is closely related to the high order of Taylor expansion of the symmetry energy, it seems difficult to constrain the $K_{sym}$ by properties of neutron-rich nuclei or matter at ground state. Probing the $K_{sym}$ by neutron star properties is recommendable but uncertainties need to be considered \cite{20191,20192,20193}.

\section{Conclusions}

To summarize, the value and the slope of nuclear symmetry energy at saturation density have been studied extensively and nowadays there are narrow constraints. There is an
interplay between the slope L and the curvature $K_{sym}$ of nuclear symmetry energy.
There is a strong correlation between the neutron-
star tidal deformability and $K_{sym}$. The curvature of nuclear symmetry energy is also closely related to the incompressibility coefficient of neutron-rich nuclear matter.
The curvature of the symmetry energy at saturation density is nowadays still largely uncertain.
It is found that in the semi-central Au+Au reaction at 400 MeV/nucleon the difference of neutron and proton elliptic flows and the squeezed-out free neutron to proton ratio at high transverse momenta are both sensitive to the curvature of the symmetry energy. At 600 MeV/nucleon the squeezed-out free neutron to proton ratio is more sensitive to the curvature of the symmetry energy. High $p_{t}$ squeezed-out n/p ratio in semi-central heavy-ion collisions seems to be one of the most potential observable to constrain the $K_{sym}$ at saturation density.

\section{Acknowledgments}

This work is supported in part by the National Natural Science Foundation of China under Grant Nos. 11775275, 11435014.

\end{document}